\documentclass[a4paper]{jpconf}
\usepackage{graphicx}
\begin{document}
\title{Quarkyonic Matter and Chiral Spirals}

\author{Toru Kojo}

\address{RIKEN BNL Research Center, Brookhaven National
  Laboratory, Upton, NY-11973, USA}

\address{torujj@quark.phy.bnl.gov}

\def\aip#1#2#3{AIP Conf. Proc. #1 (#3) #2}
\def\anp#1#2#3{Annals Phys. #1 (#3) #2}
\def\arnps#1#2#3{Ann.\ Rev.\ Nucl.\ Part.\ Sci.\ #1 (#3) #2}
\def\atmp#1#2#3{Adv. Theor. Math. Phys. #1 (#3) #2}
\def\appb#1#2#3{Acta Phys.\ Polon.\  B #1, (#3) #2}
\def\cmp#1#2#3{Comm. Math. Phys. #1 (#3) #2}
\def\cnpp#1#2#3{Comments Nucl.\Part.\Phys.\ #1 (#3) #2}
\def\cnpp#1#2#3{Comments Nucl. Part. Phys. #1 (#3) #2}
\def\epja#1#2#3{Eur. Phys. J. A #1 (#3) #2}
\def\epjc#1#2#3{Eur. Phys. J. C #1 (#3) #2}
\def\ibid#1#2#3{{\it ibid.} #1 (#3) #2}
\def\ijma#1#2#3{Intl. Jour. Mod. Phys. A #1 (#3) #2}
\def\ijmpe#1#2#3{Intl. Jour. Mod. Phys. E #1 (#3) #2}
\def\ijtp#1#2#3{Intl. Jour. Theor. Phys. A #1 (#3) #2}
\def\jhep#1#2#3{J. High Energy Phys. #2 (#3) #1}
\def\jkps#1#2#3{J. Korean Phys. Soc. #1 (#3) #2}
\def\jmp#1#2#3{Jour. Math. Phys. #1 (#3) #2}
\def\npa#1#2#3{Nucl. Phys. A #1 (#3) #2}
\def\npb#1#2#3{Nucl. Phys. B #1 (#3) #2}
\def\npsb#1#2#3{Nucl. Phys. Proc. Suppl. B #1 (#3) #2}
\def\pan#1#2#3{Phys. Atom. Nucl. #1 (#3) #2}
\def\plb#1#2#3{Phys. Lett. B #1 (#3) #2}
\def\prc#1#2#3{Phys. Rev. C #1 (#3) #2}
\def\prd#1#2#3{Phys. Rev. D #1 (#3) #2}
\def\prl#1#2#3{Phys. Rev. Lett. #1 (#3) #2}
\def\phr#1#2#3{Phys. Rep. #1 (#3) #2}
\def\ppnp#1#2#3{Prog. Part. Nucl. Phys. #1 (#3) #2}
\def\ptp#1#2#3{Prog. Theor. Phys. #1 (#3) #2}
\def\ptps#1#2#3{Prog. Theor. Phys. Supp. #1 (#3) #2}
\def\rpp#1#2#3{Rep. Prog. Phys. #1 (#3) #2}
\def\rmp#1#2#3{Rev. Mod. Phys. #1 (#3) #2}

\newcommand{\Slash}[1]{{\ooalign{\hfil/\hfil\crcr$#1$}}}
\newcommand{\Nc}{N_{\rm c}}
\newcommand{\Nf}{N_{\rm f}}
\newcommand{\muq}{\mu_q}
\newcommand{\lqcd}{\Lambda_{\rm QCD}}
\newcommand{\np}{\ :\!\!}
\newcommand{\pn}{\!\!:\ }
\newcommand{\nM}{{\cal N}_M}
\newcommand{\nm}{{\cal N}_{m_1}}
\newcommand{\nmf}{{\cal N}_{m_f}}
\newcommand{\nmc}{{\cal N}_{m_c}}
\newcommand{\PhiPE}{\Phi^\text{PE}}
\newcommand{\gammaFiveMassive}{\tilde{\gamma}_{5}}
\newcommand{\gammaZeroMassive}{\tilde{\gamma}^{0}}
\newcommand{\gammaZMassive}{\tilde{\gamma}^{z}}
\newcommand{\red}{\color{red}}

\begin{abstract}
The nuclear matter, deconfined quark matter, 
and Quarkyonic matter in low temperature region
are classified based on the $1/\Nc$ expansion.
The chiral symmetry in the Quarkyonic matter is investigated
by taking into account condensations
of chiral particle-hole pairs.
It is argued that
the chiral symmetry and parity are locally violated
by the formation of chiral spirals,
$\langle \bar{\psi} \exp(2 i \muq z \gamma^0\gamma^z ) \psi \rangle$.
An extension to multiple chiral spirals is also 
briefly discussed.
\end{abstract}
\section{Quarkyonic Matter}
Quantum Chromodynamics (QCD) in high baryon density
and low temperature
has attracted continuous interests \cite{intro}.
Recently large $\Nc$ \cite{thoofta} arguments
raised conceptual questions in dense QCD \cite{mclerran}.
It turns out that a transition scale from 
the nuclear to the quark matter ($\muq \sim \lqcd$) 
and that of deconfinement ($\muq \sim \sqrt{\Nc}\lqcd$)
are, {\it at least conceptually}, different.
The phase where a quark density is sufficiently 
high to form the Fermi sea
nevertheless excitations are confined,
is called {\it Quarkyonic} phase.

To characterize the Quarkyonic matter, below
we will first distinguish
the nuclear matter and quark matter,
and then classify the quark matter 
into the deconfined quark matter
and the Quarkyonic matter.
The $1/\Nc$ expansion will be used for theoretical
classifications.
The following arguments based on the $1/\Nc$ expansion
are only approximate,
nevertheless may be useful,
in a similar way as the chiral limit for the chiral transition,
or the heavy quark limit
for the confinement-deconfinement transition.

The nuclear matter
starts to appear slightly below the threshold of
nucleons, $\muq = M_N/\Nc \sim \lqcd$,
and changes into the quark matter
within a small variation $\sim 1/\Nc^2$ in a quark 
chemical potential.
This picture is based on the following arguments.
A change in $\muq$
requires a large change of the nucleon Fermi momentum $k_F$,
since $k_F$ is divided by a large mass
in kinetic energy $\sim k_F^2/M$.
Accordingly a number density of nucleons $\sim k_{F}^3$ 
increases rapidly, and nucleons start to overlap,
then strong nuclear forces in short distance 
becomes relevant.
In such a region,
it is natural to change effective degrees 
of freedom from nucleons
to quarks for which interactions are weaker in shorter distance.
Here quarks need not belong to particular nucleons.
A 1st order like rapid growth of a quark density
characterizes transitions from the hadronic-nuclear 
to the quark matter.
The large $\Nc$ limit provides a clear
distinction between nuclear and quark matters.

Now let us discuss properties of the quark Fermi sea.
Quarks deep inside of the Fermi sea are perturbative,
since Pauli blocking prevents them from being
affected by small momentum transfer processes.
Such quarks share a large fraction of the Fermi sea,
so perturbative estimates should reasonably work for
bulk quantities such as pressure to which all quarks contribute.
On the other hand, small momentum transfer processes
can affect quarks near the Fermi surface.
Although such quarks have large momenta $\sim \muq$,
they find comoving quarks with which
they exchange small momenta,
generating nonperturbative phenomena.
They play deterministic roles for the phase structure
(since perturbative contributions are common
to different phases), 
transport processes, which are sensitive to
excitation properties near the Fermi surface.

Now a question is how such a soft interaction looks 
like in finite density.
In the QCD vacuum, quantum fluctuations of gluons
provide the confinement,
while those of quarks screen
gluons reducing confining effects.
The strength of fluctuations can be roughly characterized
by a number of degrees of freedom: 
$O(\Nc^2)$ for gluons, and $O(\Nc)$ for quarks.
In finite density,
an allowed phase space for 
low momentum $q\bar{q}$ excitations increases,
so do screening effects.
The perturbative estimate of the screening scale is
$\sim g_s^2 \mu^2 \sim \mu^2/\Nc$,
which becomes comparable with $\lqcd$
when $\muq \sim \sqrt{\Nc}\lqcd$.
This is the scale where the deconfinement takes place.
The estimate provided here may be minimal, since
any energy gaps of quarks are not taken into account:
Mass gaps from Lorentz vector self-energy (chiral symmetric)
and possible chiral symmetry breaking
would suppress $q\bar{q}$ excitations 
with reducing screening effects.

It might sound strange to speak about the
confinement in the region where quarks are already 
released from nucleons.
To draw a qualitative picture,
we have to consider quarks in terms of quantum wavefunctions.
As a total, 
quarks occupy states in a color singlet way, 
forming a color white background.
A quark excitation in such a background
inevitablly accompanies a colored quark-hole,
and they are confined forming a mesonic state.
Glueballs are also confined,
and their spectra are discrete
without multi-gluon continuum.
A correction to this simple argument
comes from colored fluctuations in a color white background,
which are already addressed in terms of screening effects.

These arguments can be tested by studying
1+1 dimensional QCD which is a confining model.
In finite density, the strength of the confinement is unchanged
irrespective of how closely quarks are packed
within one spatial dimensional line.
This is because screening effects in finite density are always 
same as those in vacuum, 
due to the same allowed phase space
for $q\bar{q}$ fluctuations \cite{Kojo:2009ha}.

Finally it should be emphasized that
we have not used a notion of chiral symmetry breaking/restoration
to define the Quarkyonic matter.
The Quarkyonic matter is solely defined by
the presence of the quark Fermi sea and
confined excitations.
Their consequences on the chiral symmetry are
discussed in the next section.

\section{Local violation of the Chiral Symmetry: Quarkyonic Chiral Spirals}

As $\muq$ increases, the conventional particle-antiparticle
type chiral condensate disappears
since generations of antiparticles need
large energies $\sim \muq$ to transfer
particles in the Dirac sea to the above of the Fermi sea. 
Yet the chiral symmetry can be broken by
chiral pairs of particles and holes
near the Fermi surface without costing much energy.
Chiral condensates of this sort
only employ particle-holes near the Fermi surface,
modifying only particle dispersions around there.
This is consistent with arguments on perturbative 
quarks deep inside of the Fermi sea.

For simplicity,
only typical two cases,
exciton and density wave types, will be considered below
(for detail, see \cite{Kojo:2009ha}).
In the exciton case, we pick up a particle and 
a hole from the same spatial momentum region.
Since a hole momentum must be flipped,
the pair has a total momentum $\sim 0$.
For the density wave,
we pick up a particle and a hole from the
opposite momentum region,
so the pair has a total momentum $\sim 2\mu$,
forming nonuniform condensates.
Such situations have been first discussed in \cite{dgr},
in the context of the deconfined quark matter.

Since we pick up a particle and a hole
with similar kinetic energies in both cases,
main differences in pairing energies 
come from potentials.
The confining potential should provide
a big difference.
We expect that exciton wavefunctions made of 
pairs with opposite large momenta
are widely spread in space, costing large energies
due to long strings.
For density wave pairs,
particles and holes comove together
maintaining small sizes of $\sim 1/\lqcd$.
Therefore we naively expect that confining 
forces favor density wave pairings.

We first consider the one-pair problem
picking up particle-hole 
from $p_z \sim \pm \mu,\ p_T\sim0$.
This setup will turn out to be
a good starting point to analyze multi-pair problems 
\cite{Kojo:2009hb}

For concrete arguments, 
we introduce a simple model
of the linear confinement in which 
gluon propagators take the following form
(screening effects are suppressed in large $\Nc$),
\begin{equation}
D_{44}^{A B}(k) 
= - \frac{8\pi}{C_F} \times \frac{\sigma }{(\vec{k}^2)^2} \; \delta^{A B}
\; \; \; ; \;\;\;  
D^{4 i} = D^{ij} = 0 \ , \ \ 
(C_F=\frac{\Nc^2-1}{2\Nc},\ \sigma\sim \lqcd^2)
\label{gluon_propagator}
\end{equation}
which shows a linear potential in coordinate space.
This model is inspired
by Coulomb gauge analyses \cite{zwanziger}.
We have omitted perturbative parts for 
the sake of simplicity 
\cite{glozman}. 

To investigate condensations,
we analyze self-consistent equations
by reducing 3+1 dimensional equations to 1+1 dimensional ones.
(The essence was already discussed in \cite{son}).
Below we illustrate this for the Schwinger-Dyson equation
for the quark self-energy. 
It looks like
\begin{equation}
\Slash{\Sigma}(p) +\Sigma_m(p)
= - \int \frac{d^4 k}{(2\pi)^4} \;
 (\gamma_4 t_A) \; S(k;\Sigma) \; (\gamma_4 t_B)
\; D^{AB}_{44}(p-k)\;,
\label{SD1}
\end{equation}
where the quark propagator is dressed and
depends on $\Sigma(p)$.
We will investigate the self-energy
for quarks near the Fermi surface, with
$p_z \sim \mu$ and $p_T \sim 0$.

Three points are relevant
for the dimensional reduction:
(I) Our gluon propagator behaves as $1/(\vec{q}^2)^2$,
so dominant contributions of the integral (\ref{SD1})
sharply concentrates on the small domain.
(II) Different Dirac structures
$S(k) 
= \gamma_4 S_4(k) + \gamma_z S_z(k) + \vec{\gamma}_T \cdot \vec{S}_T(k)
+ S_m(k)$, give
the ratio of $|S_T|/S_z \sim k_T/k_z \sim \lqcd/\mu$,
so we can drop off $S_T$ and $\Sigma_T$ in the leading order of 
$\lqcd/\mu$.
(III) The quark energy is sensitive to the change of 
$k_z$ while not to that of $k_T$.
Changes in $k_T$ is along the constant 
energy surface which looks flat for $\lqcd/\mu \ll 1$,
without changing the quark energy a lot. 

Because of the restricted integral region and
the insensitivity of the quark propagator to $k_T$ variable,
we can set $\vec{k}_T \simeq \vec{0}_T$ 
and factorize the integral equation:
\begin{equation}
\int dk_4 dk_z d^2\vec{k}_T 
\; S(k_4, k_z, \vec{k}_T) \; D_{44}(p-k)\;
\rightarrow 
\int dk_4 dk_z  
\; S(k_4, k_z, \vec{0}_T) 
\int d\vec{k}_T \; D_{44}(p-k),
\label{SD1}
\end{equation}
for which we can carry out the $\vec{k}_T$ integral.
The smearing of gluon propagator yields a
1+1 dimensional confining propagator, 
$\sim 1/(p_z - k_z)^2$.
The reduced equation then becomes
\begin{equation}
\bar{\Slash{\Sigma} } (p_4, p_z, \vec{0}_{\perp}) 
+ \Sigma_m (p_4, p_z, \vec{0}_{\perp})
\simeq
\frac{\Nc g^{2}_{{\rm 2D}} }{2} \; \int \frac{dk_4 \, dk_z}{(2\pi)^2}\;
\gamma_4 \; \bar{S}(k_4,k_z, \vec{0}_{\perp}) \; \gamma_4 \;
\frac{1}{\left(k_z-p_z\right)^2} \ ,
\label{SD5}
\end{equation}
where $\Nc g^2_{\rm 2D} = 4\sigma$
and we denote $\bar{\Sigma}, \bar{S}$ 
for components other than transverse ones.          
The resulting equation is nothing but
the Schwinger-Dyson equation of
't Hooft model in axial gauge $A_z=0$.
The same approximations are applicable to 
the Bethe-Salpeter equation.

Since self-consistent equations
for our one-pair problem take the same form
as those of 't Hooft model \cite{thooftb},
we can borrow results in the existing literatures \cite{bringoltz}.
Only nontrivial issues are relationships between
3+1 and 1+1 dimensional operators and condensation channels.
In particular, concepts of spins are absent in
one spatial dimension.
In our dimensional reduction,
the tranverse part of the quark propagators are suppressed. 
This implies that $\vec{\gamma}_T$ terms are dropped off,
so that there is no spin mixing term once
we quantize spins along the directions of moving particles.
Introducing projection operators for
moving directions, we can write spin multiplets
\cite{son},
\begin{equation}
\psi_{\pm} = \frac{1 \pm \gamma^0 \gamma^z}{2}\psi \; ,
\;\;\;\;\;\;
\varphi_\uparrow 
= \left[  
\begin{array}{cc}
\varphi_{\uparrow +}\\  
\varphi_{\uparrow -}
\end{array}
\right] 
=
\left[  
\begin{array}{cc}
\psi_{R+}\\  
\psi_{L-} 
\end{array}
\right] 
\;\;\; ; \;\;\;
\varphi_\downarrow 
= 
\left[  
\begin{array}{cc}
\varphi_{\downarrow +}\\  
\varphi_{\downarrow -}
\end{array}
\right] 
=
\left[ 
\begin{array}{cc}
\psi_{L+} \\
\psi_{R-} 
\end{array}
\right] \; ,
\end{equation}
where indices ($+,-$) for $+z$,$-z$ moving particles.
Let us introduce $1+1$ dimensional "flavor", 
$\Phi^T \equiv (\varphi_\uparrow, \varphi_\downarrow)$,
and Dirac matrices,
$\Gamma^0 = \sigma^1, \;
\Gamma^z = - i \sigma^2, \;
\Gamma^5 = \sigma^3$.
Then 3+1 dimensional quark bilinears without 
spin mixings are mapped onto 
1+1 dimensional "flavor" {\it singlet} operators
\cite{Kojo:2009ha,son},
\begin{eqnarray}
\bar{\psi} \psi \rightarrow \bar{\Phi} \Phi, \;\;\;\;\;
\bar{\psi} \gamma^0 \psi \rightarrow \bar{\Phi} \Gamma^0 \Phi, \;\;\;\;\;
\bar{\psi} \gamma^z \psi \rightarrow \bar{\Phi} \Gamma^z \Phi, \;\;\;\;\;
\bar{\psi} \gamma^0 \gamma^z \psi
  \rightarrow \bar{\Phi} \Gamma^5 \Phi.
\end{eqnarray}
The last relation reflects the fact that
1+1 dimensional $\Gamma^5$ characterizes
moving directions.
All other quark bilinears include spin mixings,
so they are "flavor" {\it non-singlet} in 1+1 dimensions.
They do not show condensations in the 't Hooft model.

The 't Hooft model in finite density
is known to show the chiral spirals
rotating in chiral space as 
$\langle \bar{\Phi} e^{2i \muq z \Gamma^5} \Phi \rangle$.
Mapping this, we get the chiral spiral in 3+1 dimensions,
\begin{eqnarray}
\langle \bar{\psi} \psi \rangle = \Delta \cos(2\mu z),\ \ 
\langle \bar{\psi} \gamma^0 \gamma^z \psi \rangle = \Delta \sin(2\mu z).
\end{eqnarray}
An origin of this chiral rotation is that
$\bar{\psi}_{+}\psi_{-}$ and $\bar{\psi}_{-}\psi_{+}$
moving to opposite directions each other,
thus have opposite phase $e^{\pm 2i\mu z}$; 
this phase mismatch makes
$\langle \bar{\psi} \gamma^0 \gamma^z \psi  \rangle
= \langle \bar{\psi}_{+}\psi_{-} \rangle
- \langle \bar{\psi}_{-}\psi_{+} \rangle$
nonvanishing.
So the chiral density wave accompanies
its partner, forming spiral structures.

It is clear that 
a spatial average of chiral condensates 
vanishes because of spatial modulations.
So the symmetry apparently looks restored
for probes with large wavelength.
Note also that our chiral tensor type 
condensate locally breaks parity.

Finally we briefly mention
an extension of the present results to
those of multi-pairs.
This problem was investigated in \cite{Kojo:2009hb}.
A key observation is that
in confining models,
chiral spirals evolved in different spatial directions
interact only weakly.
Therefore our results for one-patch problem 
may be applied with slight modifications.
Such a picture indicates that
a number of directions of chiral spiral formations
increases as $\muq$,
with a series of phase transitions.

\section{Summary and Outlook}

In the Quarkyonic phase,
excitations are confined despite of 
a large quark number density.
They drive the formation of 
the chiral spirals with large mass gaps $\sim \lqcd$.
This is in sharp contrast to the situation in the deconfined quark matter 
where gaps are too small to overtake screening effects \cite{son}.
So chiral spirals can be regarded 
as good signatures of the Quarkyonic phase.
Transport processes in stars are good candidates 
to see consequences of chiral spirals
through characteristic Nambu-Goldstone modes.
A local parity violation may also provide
important implications through neutrino physics.
These issues deserve further investigations.

\ack
The author acknowledges his collaborators, 
Y. Hidaka,
L. McLerran, R.D. Pisarski, and A.M. Tsvelik
with whom the present ideas have been developed.
This research is supported under 
DOE Contract No. DE-AC02-98CH10886
and Special Posdoctoral Research Program of RIKEN.
He also thanks organizers and participants
of the workshop Hot Quark 10.



\end{document}